\documentstyle[bibnorm]{lamuphys}
\def\be{\begin{equation}}       \def\eq{\begin{equation}}
\def\ee{\end{equation}}         \def\eqe{\end{equation}}

\def\bea{\begin{eqnarray}}      \def\eqa{\begin{eqnarray}}
\def\ena{\end{eqnarray}}        \def\eea{\end{eqnarray}}
                                \def\eqae{\end{eqnarray}}

\def\g0{g_{(0)}}
\def\gi{g_{(0)}^{-1}}
\def\Tr{{\rm Tr}}
\def\pa{\partial}
\def\G{\Gamma}

\begin{document}


\title{Holography and the Weyl anomaly}

\author{M{\aa}ns\,Henningson\inst{1}\and 
Kostas\,Skenderis\inst{2}}

\institute{Institute of Theoretical Physics, \\
Chalmers University of Technology,\\
S-412 96 G\"oteborg, Sweden\\
{\tt mans@fy.chalmers.se}
\and
Spinoza Institute, University of Utrecht, \\
Leuvenlaan 4, 3584 CE Utrecht, The Netherlands\\
{\tt k.skenderis@phys.uu.nl}}

\maketitle

\begin{abstract}
We review our calculation of the Weyl anomaly for $d$-dimensional conformal 
field theories that have a description in terms of a $(d + 1)$-dimensional 
gravity theory.   
\end{abstract}
\section{Introduction}
In the last few months, we have learned that a theory containing gravity and 
defined on an open $(d + 1)$-manifold $X$ can, in some cases, be equivalent to 
a $d$-dimensional conformal field theory defined on the boundary $M$ of $X$ 
\cite{ads/cft1}, \cite{ads/cft2}, \cite{ads/cft3}. The manifold 
$X$ has a causal structure similar to that of $(d + 1)$-dimensional anti-de 
Sitter space, i.e. the boundary $M$ at conformal infinity is timelike. This 
means that the space $X$ is not globally hyperbolic, i.e. it does not admit 
any Cauchy surfaces, since new information may come in through the 
boundary $M$. The action functional $S[\phi]$ or the equations of motion for 
the gravity theory on $X$ therefore do not completely determine the dynamics 
and must be supplemented by appropriate boundary data $\phi_{(0)}$ on $M$ for 
each field $\phi$ on $X$. The partition function is then a functional of these
boundary data:
$$
Z_{grav}[\phi_{(0)}] = \int_{\phi_{(0)}} {\cal D} \phi \exp ( - S[\phi]) ,
$$
where the subscript on the integration sign indicates that the functional 
integral is over field configurations $\phi$ that satisfy the boundary 
condition given by $\phi_{(0)}$. The precise relationship between this 
gravity theory on $X$ and the conformal field theory on the boundary $M$ can 
now be described as follows: There is a one-to-one correspondence between the 
fields $\phi$ on $X$ and the primary operators ${\cal O}$ on $M$. The set of 
correlation functions of the latter are conveniently summarized by a 
generating functional:
$$
Z_{CFT}[\phi_{(0)}] = \Bigl\langle \exp \int_M d^d x {\cal O} \phi_{(0)} 
\Bigr\rangle ,
$$
where $\phi_{(0)}$ is now regarded as a formal expansion parameter. The 
partition function of the gravity theory on $X$ and the generating functional 
on $M$ are then equal, regarded as functionals of $\phi_{(0)}$:
$$
Z_{grav}[\phi_{(0)}] = Z_{CFT}[\phi_{(0)}] .
$$
This relationship is particularly useful when the gravity theory is weakly 
coupled. We can then calculate the partition function at tree level, where it 
is given by the exponential of the action functional evaluated for a field 
configuration $\phi^{cl}(\phi_{(0)})$ that solves the classical equations of 
motion and obeys the boundary conditions given by $\phi_{(0)}$:
$$
Z_{grav}^{tree}[\phi_{(0)}] = \exp \left( - S[\phi^{cl}(\phi_{(0)})] \right).
$$ 
 
A field of particular importance in a theory containing gravity is of course 
the metric $G_{\mu \nu}$. The corresponding operator in the boundary conformal
field theory is the stress-energy tensor $T_{ij}$. The boundary data for the 
metric $G_{\mu \nu}$ is {\it not} a boundary metric $g_{(0)\ ij}$, but 
only a conformal structure $[g_{(0)}]$. (This is defined as an equivalence 
class of boundary metrics when two metrics that differ by a local rescaling 
are considered equivalent, i.e. $g_{(0)} \sim \exp 2 \sigma (x) g_{(0)}$ for 
an arbitrary positive function $\sigma (x)$.) Indeed, the metric $G_{\mu \nu}$
on $X$ has a second order pole at the boundary $M$ and thus does not 
induce a metric there. To get a finite metric on $M$, we could multiply 
$G_{\mu \nu}$ by $f^2$, where $f$ is some positive function on $X$ with a 
simple zero on $M$ and then restrict to $M$. The freedom to choose $f$ means 
that only the conformal equivalence class of the metric so obtained is 
unambiguously defined. Conversely, a choice of conformal structure 
$[g_{(0)}]$ on $M$ together with Einstein's equations on $X$ suffice 
to determine an (up to diffeomorphisms) unique metric $G_{\mu \nu}$ on $X$ 
\cite{GrahamLee}. We 
thus naturally get a {\it conformal} field theory on $M$. Indeed, to consider 
a more general theory on $M$, we  would need a background metric. However, as 
we have seen, only the conformal structure of $M$ is well-defined, so only 
conformal field theories are meaningful on $M$.

In the following, we will consider the effective action $W_{CFT}[g_{(0)}] = - 
\log Z_{CFT}[g_{(0)}]$ of the theory on $M$. This can be regarded as minus 
the logarithm of the zero-point function or partition function of the theory. 
(It is a zero-point function in the sense that there are no insertions of any 
operators). A priori, this is a functional 
of the metric $g_{(0)}$ on $M$, but by conformal invariance, it should 
actually only depend on the conformal equivalence class $[g_{(0)}]$. However, 
this invariance is sometimes broken by a conformal (or Weyl) anomaly. 
This means 
that $W_{CFT}[g_{(0)}]$ is not invariant under a conformal rescaling $\delta 
g_{(0)} = 2 \delta \sigma g_{(0)}$ of the metric, but transforms as
$$
\delta W_{CFT}[g_{(0)}] = \int_M d^d x \sqrt{\det g_{(0)}} {\cal A} \delta 
\sigma,
$$
where ${\cal A}$ is the anomaly. On general grounds, the gravitational
part of the Weyl anomaly vanishes when 
the dimension $d$ of $M$ is odd. When $d$ is even, it is of the form
$$
{\cal A} = E + I + D_i J^i ,
$$
where $E$ is proportional to the $d$-dimensional Euler density, and $I$ is a 
conformal invariant \cite{BonoraPastiBregola}
\cite{DeserSchwimmer}. The total derivative term $D_i J^i$ is 
irrelevant, since it can be canceled by adding a local counterterm to the 
effective action. There is of course a unique Euler density in  every even 
dimension $d$. The number of linearly independent conformal invariants grows 
with $d$, though. There are no such invariants for $d = 2$, one (the square of
the Weyl tensor) for $d = 4$, three for $d = 6$, etcetera. We will outline how
to calculate the conformal anomaly for $d$-dimensional conformal field 
theories that have a description in terms of a $(d + 1)$-dimensional gravity 
theory as described above. Technical details (and some
intermediate results not presented in \cite{HenningsonSkenderis})
are relegated to the appendix.

\section{The calculation}
According to the recipe above, $W_{CFT}[g_{(0)}]$ is given by minus the 
logarithm of the partition function of the gravity theory with a certain 
conformal structure $[g_{(0)}]$ induced by the metric $G_{\mu \nu}$ on the 
boundary. All other fields of the gravity theory should vanish on the 
boundary $M$, since we are not inserting any operators in the conformal field 
theory correlation function. If a tree-level computation is justified, so 
that we only need to consider field configurations on $X$ that solve the 
classical equations of motion, this means that all fields expect the metric 
vanish everywhere on $X$. The theory in bulk is then reduced to pure gravity 
described by the Einstein-Hilbert action plus a cosmological constant term:
\be \label{act1}
S_{bulk} = {1 \over 16 \pi G_N} \int_X d^{d + 1} x \sqrt{\det G_{\mu \nu}} 
\left( R + 2 \Lambda \right) .
\ee
On a manifold with boundary, we also have the term
\be \label{act2}
S_{boundary} = {1 \over 16 \pi G_N} \int_M d^d x 
\sqrt{\det \tilde{g}_{ij}} 
2 K ,
\ee
where $\tilde{g}_{ij}$ is the metric on the boundary and $K$ is the 
trace of the second fundamental form. This boundary term
is necessary in order to get an action that depends only on 
first derivatives of the metric\cite{GibbonsHawking}. 
As described in the 
first section, a choice of conformal structure $[g_{(0)}]$ on the boundary 
$M$ determines a unique  metric $G_{\mu \nu}$ in the bulk of $X$ that solves 
the equations of motion
\begin{equation} \label{Einst}
R_{\mu \nu} - {1 \over 2} R G_{\mu \nu} = \Lambda G_{\mu \nu} . 
\end{equation}
(This is of course true only up to diffeomorphisms). However, the bulk 
action diverges when evaluated for such a field configuration because of the 
second order pole in $G_{\mu \nu}$ on the boundary. Furthermore, the boundary 
terms in the action are ill-defined, since $G_{\mu \nu}$ does not induce a 
finite metric $\tilde{g}_{ij}$ on $M$.

To regulate the theory in a manner consistent with general covariance, we 
pick a specific representative $g_{(0)}$ of the boundary conformal 
structure $[g_{(0)}]$. This determines a distinguished coordinate system 
$(\rho, x^i)$, in which the metric on $X$ takes the form 
\cite{FeffermanGraham}
\begin{equation} \label{Gdef}
G_{\mu \nu} d x^\mu d x^\nu = {l^2 \over 4} \rho^{-2} d \rho d \rho + 
\rho^{-1} g_{ij} d x^i d x^j .
\end{equation}
Here the tensor $g$ has the limit $g_{(0)}$ as one approaches the boundary 
represented by $\rho = 0$. The length scale $l$ is related to the cosmological 
constant $\Lambda$. Einstein's equations for $G_{\mu \nu}$ can then be solved 
order by order in $\rho$ with the result that $g$ is of the form
\be \label{even}
g = g_{(0)} + \rho g_{(2)} + \ldots + \rho^{d/2} g_{(d)} + \rho^{d/2} \log 
\rho \, h_{(d)} + {\cal O}(\rho^{d/2 + 1}) . 
\ee
The regularization procedure now amounts to restricting the bulk integral in 
the action to the domain $\rho > \epsilon$ for some cutoff $\epsilon > 0$ and 
evaluating the boundary integrals at $\rho = \epsilon$. The regulated action 
evaluated for this field configuration is then of the form
\bea
W[g_{(0)}] & = & {1 \over 16 \pi G_N} \int d^d x \sqrt{\det g_{(0)}} \left( 
\epsilon^{-d/2} a_{(0)} + \epsilon^{-d/2 + 1} a_{(2)} + \ldots + \epsilon^{-1} 
a_{(d - 2)} - \log \epsilon a_{(d)} \right) \cr \nonumber \\
&& + W_{fin}[g_{(0)}] . \label{regact}
\eea

The coefficients $a_{(0)}, a_{(2)}, \ldots, a_{(d)}$ are all given by 
covariant expressions in $g_{(0)}$ and its curvature tensor $R^i{}_{jkl}$
(see appendix). 
The divergences as $\epsilon$ goes to zero can thus be canceled by adding 
local counterterms to the action, so that we are left with a finite effective 
action $W_{fin}[g_{(0)}]$. To find the variation of $W_{fin}[g_{(0)}]$ under 
a conformal rescaling of the boundary metric $g_{(0)}$, we note that the 
regulated action $W[g_{(0)}]$ is invariant under the combined transformation 
$\delta g_{(0)} = 2 \delta \sigma g_{(0)}$ and $\delta \epsilon = 2 \delta 
\sigma \epsilon$ for a {\it constant} parameter $\delta \sigma$. The terms 
proportional to negative powers of $\epsilon$ are separately invariant, so 
the variation of $W_{fin}[g_{(0)}]$ must therefore equal minus the variation 
of the logarithmically divergent term. The latter is given by
$$
{\cal A} \delta \sigma = {1 \over 16 \pi G_N} (- 2 a_{(d)}) \delta \sigma ,
$$
since $\log \epsilon$ transforms with a shift whereas $a_{(d)}$ itself is 
invariant. Although this formula was derived under the assumption that 
$\delta \sigma$ is a constant, it follows from the general form of the 
conformal anomaly that it can be applied also for a non-constant $\delta 
\sigma$.

\section{The results}
The discussion in the preceding section shows that, up to a constant, the 
conformal anomaly of a theory related to a gravity theory only depends on the 
space-time dimension $d$. We will now evaluate the anomaly explicitly for the 
physically relevant cases $d = 2, 4, 6$.

\subsection{$d = 2$ and the asymptotic symmetry algebra of $adS_3$}
Here we get
$$
{\cal A} = - {c \over 24 \pi} R 
$$
with the central charge
$$
c = {3 l \over 2 G_N} .
$$
This agrees with a computation based on the asymptotic symmetry algebra of 
three-dimensional anti-de Sitter space \cite{BrownHenneaux}!

\subsection{$d = 4$ and ${\cal N} = 4$ super Yang-Mills theory}
Here we get
$$
{\cal A} = - {2 l^3 \over 16 \pi G_N} \left( - {1 \over 8} R^{ij} R_{ij} + 
{1 \over 24} R^2 \right) .
$$
Inserting for example the values of $l$ and $G_N$ appropriate for the $adS_5 
\times S^5$ geometry of $N$ coincident $D3$-branes in type IIB string theory, 
we get $N^2$ ($\sim$ dim $SU(N)$) times the conformal anomaly of an ${\cal N} 
= 4$ supermultiplet (one vector, four chiral spinors and six scalars) 
\cite{Duff}! 
The agreement between our strong coupling calculation and this weak coupling 
result indicates that there is a non-renormalization theorem for the conformal 
anomaly. 

\subsection{$d = 6$ and tensionless strings}
Here we get
\bea
{\cal A}& = & 
- {2 l^5 \over 16 \pi G_N} \left( - {1 \over 128} R R^{ij} R_{ij} 
+ {3 \over 3200} R^3 + {1 \over 64} R^{ij} R^{kl} R_{ijkl} \right. \cr
\nonumber \\
&& + \left. {1 \over 320} R^{ij} D_i D_j R - {1 \over 128} R^{ij} D^k D_k 
R_{ij} + {1 \over 1280} R D^i D_i R \right) . \nonumber
\eea
Inserting the values of $l$ and $G_N$ corresponding to the $adS_7 \times S^4$ 
geometry of $N$ coincident five-branes in $M$-theory, we find that ${\cal A}$ 
is proportional to $N^3$ ($>>$ dim $GL(N)$). This agrees with considerations 
based on the entropy of the brane system \cite{GubserKlebanov},
\cite{KlebanovTseytlin}. This provides information 
about the mysterious tensionless string theory that appears when $M5$-branes 
coincide.

\appendix
\section{Appendix}

In this appendix we present some technical details of the calculation 
of the conformal anomaly.

Einstein's equation (\ref{Einst}) for $G$ given in (\ref{Gdef}) amount to
\bea
\rho \left(2 g^{\prime\prime} - 2 g^\prime g^{-1} g^\prime + \Tr
(g^{-1} g^\prime) g^\prime \right) + l^2 {\rm Ric} (g) - (d - 2)
g^\prime - \Tr (g^{-1} g^\prime) g & = & 0 \cr
(g^{-1})^{jk} \left(\nabla_i g_{jk}^\prime - \nabla_k g_{ij}^\prime
\right) & = & 0 \cr
\Tr (g^{-1} g^{\prime\prime}) - \frac{1}{2} \Tr (g^{-1} g^\prime
g^{-1}
g^\prime) & = & 0 , \label{eqn}
\eea
where differentiation with respect to $\rho$ is denoted with a prime,
$\nabla_i$ is the covariant derivative constructed from the metric
$g$
and ${\rm Ric} (g)$ is the Ricci tensor\footnote{
Our conventions are as follows
$R_{ijk}{}^l=\pa_i \G_{jk}{}^l + \G_{ip}{}^l \G_{jk}{}^p - i
\leftrightarrow j$ and $R_{ij}=R_{ikj}{}^k$.} of $g$.
We solve these equations iteratively in $\rho$.

The $g_{(k)}$ for $k < d$ in (\ref{even}) are given by
\bea
g_{(2)}{}_{ij} & = & \frac{1}{d - 2} \left( R_{ij} - \frac{1}{2 (d - 1)} 
R \g0{}_{ij} \right) \cr
g_{(4)}{}_{ij} & = & \frac{1}{d - 4} \left( - \frac{1}{8 (d - 1)} D_i
D_j R + \frac{1}{4 (d - 2)} D_k D^k R_{ij} \right . \cr
& & - \frac{1}{8 (d - 1) (d - 2)} D_k D^k R \g0{}_{ij} - \frac{1}{2 (d - 2)}
R^{kl} R_{ikjl} \cr
& & + \frac{d - 4}{2 (d - 2)^2} R_i{}^k R_{kj} + \frac{1}{(d - 1)(d -
2)^2} R R_{ij} \cr
& & \left. + \frac{1}{4 (d - 2)^2} R^{kl} R_{kl} \g0{}_{ij} - \frac{3 d}{16
(d - 1)^2 (d - 2)^2} R^2 \g0{}_{ij} \right) .
\eea
Although the above formulas for $g_{(k)}{}_{ij}$ do not make sense
for $d = k$, their traces are smooth in the $d$ goes to $k$ limit and
in fact give the correct values for 
${\rm Tr}(g_{(0)}^{-1}g_{(k)})$ for $k \leq d$.
Actually the easiest way to calculate the traces is to use the last equation 
in (\ref{eqn}). By differentiating this equation with respect to $\rho$ 
and then setting $\rho$ to zero one obtains
\bea
&&\Tr \left(\g0^{-1} g_{(4)}\right) = {1 \over 4} 
\Tr \left((\g0^{-1} g_{(2)})^2 \right) \nonumber \\
&& \Tr \left(\g0^{-1} g_{(6)}\right) = {2 \over 3} 
\Tr \left(\g0^{-1} g_{(2)} \g0^{-1} g_{(4)}\right)
-{1 \over 6} \Tr \left((\g0^{-1} g_{(2)})^3 \right)
\eea
The above data are sufficient in order to obtain the 
anomaly up to $d=6$. In general, to determine the Weyl anomaly 
at dimension $d$ one needs to determine the full metric up to order 
$d-2$. The trace of the metric at order $d$ can be determined 
from metric at lower orders using the last equation in (\ref{eqn}).

The explicit expression of $a_{(n)}$'s in (\ref{regact}) can be derived
by inserting the general form of the metric (\ref{Gdef}) in 
(\ref{act1}), perform the $\rho$ integration, and evaluate (\ref{act2})
at $\rho=\epsilon$. The coefficients $a_{(d)}$ (in $d$ dimensions)
of the logarithmic divergences, which are relevant for the calculation 
of the Weyl anomaly,  receive contributions only from 
the bulk integral. Their explicit form for $d=2,3,6$, 
are given by
\bea
a_{(2)}&=& l \, \Tr (\gi g_{(2)}) \nonumber \\
a_{(4)}&=& l^3 \frac{1}{2} \left([\Tr (\gi g_{(2)})]^2 
- \Tr [(\gi g_{(2)})^2]\right) \nonumber \\
a_{(6)} & = & l^5 \left( {1 \over 8} [\Tr \gi g_{(2)}]^3
-{3 \over 8} \Tr[\gi g_{(2)}] \Tr[(\gi g_{(2)})^2]\right. \nonumber \\
&& \left.+{1 \over 2} \Tr[(\gi g_{(2)})^3]
- \Tr[\gi g_{(2)} \gi g_{(4)}] \right). \nonumber
\eea

\end{document}